\documentclass[12pt]{amsart} 
\usepackage{xypic,amssymb}

\theoremstyle{definition}

\theoremstyle{remark}

\numberwithin{equation}{section}

\newcommand\op[1]{\operatorname{#1}}
\newcommand\End{\operatorname{End}}
\newcommand\Ext{\operatorname{Ext}}
\newcommand\Hom{\operatorname{Hom}}
\newcommand\iso{\kern.35em{\raise3pt\hbox{$\sim$}\kern-1.1em\to}\kern.3em}

\newcommand\F{{\mathcal F}}\newcommand\Oc{{\mathcal O}}
\newcommand\Pc{{\mathcal P}}\newcommand\M{{\mathcal M}}
\newcommand\Rcal{{\mathcal R}}
\newcommand\R{{\mathbb R}}

\newcommand\ch{\operatorname{ch}}

\newcommand\Ps{{\mathbb P}}

\newcommand\G{{\mathcal G}}

\newcommand\Hc{{\mathcal H}}

\newcommand\Ic{{\mathcal I}}

\begin{document}
\title{Comments on $N=1$ Heterotic String Vacua}
\author{ B. Andreas\thinspace$^{\mbox{\S}}$ }
\author{ D. Hern\'andez Ruip\'erez $^{\mbox{\ddag}}$}
\email{bandreas@maths.harvard.edu, ruiperez@usal.es}
\address{$^{\mbox{\S}}$\thinspace Institut f\"ur Mathematik, Humboldt Universit\"at zu
Berlin, D-10115 Berlin, Germany}
\address{$^{\mbox{\ddag}}$\thinspace Departamento de Matem\'aticas, Universidad de Salamanca, Plaza de la Merced 1-4, 37008 Salamanca, Spain}
\date{\today} 
\thanks {Research supported by DFG Schwerpunktprogramm (1096)
``Stringtheorie im Kontext von Teilchenphysik, Quantenfeldtheorie,
Quantengravitation, Kosmologie und Mathematik'', by the Spanish {\sc dges} (research project BFM2000-1315) and  by ``Junta de Castilla y Le\'on'' 
(research project SA009/01). The second author is member of  {\sc vbac} (Vector Bundles on Algebraic
Curves), which is partially supported by {\sc eager} ({\sc ec fp5}
Contract no. {\sc hprn-ct-2000-00099}) and
by {\sc edge} ({\sc ec fp5} Contract no. {\sc hprn-ct-2000-00101}).}

\thanks{Also available as preprint {\tt hep-th/0305123}. }
\begin{abstract} We analyze three aspects of $N=1$ heterotic string compactifications
on elliptically fibered Calabi-Yau threefolds: stability of vector
bundles, five-brane instanton transitions and chiral matter. First we
show that
relative Fourier-Mukai transformation preserves absolute stability.
This is
relevant for vector bundles whose spectral cover is reducible. Then we
derive an
explicit formula for the number of moduli which occur in (vertical)
five-brane instanton transitions provided a certain vanishing argument
applies. Such transitions increase the holonomy of the heterotic vector
bundle and cause gauge changing phase transitions. In an M-theory
description the transitions are associated with collisions of bulk
five-branes with one of the boundary fixed planes. In
F-theory they correspond to three-brane instanton transitions. Our
derivation relies on an index computation with data localized along the
curve which
is related to the existence of chiral matter in this class of heterotic
vacua. Finally, we
show how to compute the number of chiral matter multiplets for this
class of
vacua allowing to discuss the associated Yukawa couplings.
\end{abstract}
\maketitle 

\section{Introduction}
One of the important issues in studying string compactifications is to
understand the moduli space of four-dimensional $N=1$ string vacua.
The compactification of the heterotic string on a Calabi-Yau threefold
$X$ with a stable, holomorphic vector bundle $V$ is one way to obtain
such vacua.

In recent years, there has been tremendous progress in understanding
the class of four-dimensional $N=1$ vacua obtained by compactification
on
elliptically fibered Calabi-Yau manifolds. This class has the double
advantage
to admit an explicit description of vector bundles in terms of spectral
covers \cite{FMW} and to
allow a dual description in terms of F-theory.

To obtain a consistent heterotic string compactification on an elliptic
Calabi-Yau threefold one has to include a number of five-branes
which wrap the elliptic fibers \cite{FMW}. It has
been shown that these five-branes match precisely the number of
space-time filling three-branes necessary for tadpole cancellation in
F-theory
compactified on elliptically fibered Calabi-Yau fourfolds $Y$ \cite{FMW}.
Various aspects
of the map between the geometrical
moduli of the pair $(X,V)$ and those of $Y$ have been studied in
\cite{FMW, BJPS, DI, DIA, AC, mv, PB}. The question of which pairs
$(X,V)$ are stable under world-sheet instanton corrections \cite{SiWi} has
been
recently reconsidered in \cite{BaSe, BeaWi}.

Besides an improved understanding of this map and the discussion of
phenomenologically viable pairs $(X,V)$, one would like to understand
the behavior of $(X,V)$ at singularities. Such singularities can be
either
associated to a degeneration of $X$ or $V$. In general one expects
there new non-perturbative
effects, associated to the breakdown of world-sheet conformal field
theory.
A well known example is the small instanton singularity in heterotic
string
compactifications on a $K3$ surface \cite{WIT}.
The heterotic vector bundle degenerates to a torsion free sheaf with
singularity locus of codimension at most two  \cite{ASP, SHA}. Thus
from a mathematical point of view, the bundle fails to be local free at
a finite number of
points on the $K3$ surface. One can interpret \cite{SW} such small
instantons as five-branes whose world volumes fill the six
uncompactified
directions and intersect the $K3$ surface at these points
\cite{GHa, SW, DMM}. The
observable effects are a change in the unbroken gauge symmetry and the
number of tensor multiplets.

In passing to four-dimensional compactifications of the heterotic
string a similar picture is expected to hold. In particular the class
of elliptically
fibered Calabi-Yau threefolds allows an explicit study of such
transitions. For this class
one has three possibilities of codimension-two bundle degeneration
associated to curves in the
base, the fibers or linear combinations of both. Like in the
six-dimensional situation these degenerations can be
interpreted in terms of five-branes wrapping over these holomorphic
curves. In addition to codimension-two degenerations one expects
pointlike
bundle singularities in codimension-three which have been studied in
\cite{CDFMR}.

The codimension-two degenerations have in common that the heterotic
vector bundle is associated (via fiberwise T-duality or relative
Fourier-Mukai transformation) to
the spectral data $(C,L)$ with $C$ being a reducible spectral cover
\cite{OPP} and $L$ the
spectral line bundle. Now the generic heterotic vector bundle
associated with an irreducible
spectral cover is stable. This leads to a stability question of $V$ in
the reducible
case which we will tackle in section 3.

Now the observable effects in four dimensions depend on whether the
bundle degenerates over base or fiber curves (or linear combinations of
both). If the
bundle degenerates over a base curve one observes (after smoothing out
the singular gauge
configuration) a change in the charged matter content \cite{OPP}. This
provides evidence for
chirality changing phase transitions in four-dimensional string vacua
\cite{KASI}. As the second Chern
class gets shifted by the cohomology class of the associated curve, one
can effectively interpret such transition as five-brane instanton
transition \cite{OPP}. On the other hand, bundle degeneration associated to
fiber curves do not change the net-amount of chiral matter (the third
Chern class is left unchanged), however, they do
change the structure group of the heterotic vector bundle and thus the
unbroken gauge group in four dimensions.

Both transition types are expected to have a dual interpretation in
terms of F-theory.
In particular, one expects chirality changing transitions to be dual to
a change in the
F-theory four form flux \cite{AnC, OPP}. Gauge changing
transitions are expected to
be dual to three-brane instanton transitions \cite{BJPS}. More precisely the
number of five-branes
which dissolve in such transition are supposed to match the precise
number of three-branes
in F-theory.  One can also consider the transitions in heterotic
M-theory \cite{BDO, DLO, MPS}.
The anomaly cancellation requires to include additional five-branes in
the bulk space.
The transition is then interpreted as `collision' of a bulk five brane
with one of the boundary
fixed planes. Conditions under which the five-brane is attracted to
the boundary fixed plane are discussed in \cite{MPS}.

The geometrical moduli of the pair $(X,V)$ are given by the complex
structure and K\"ahler deformations
of $X$ and the bundle moduli given by the dimension of $H^1(X,
\End(V))$. One would like to
know how the moduli of $V$ are altered in chirality or gauge changing
phase transitions.
For chirality changing transitions the question has been studied in
\cite{BDO}. In particular, it was
shown in \cite{BDO} that the transition moduli (the difference of the moduli
of the original and transition
bundle) can be interpreted as moduli of the altered spectral cover
restricted to the lift of the
horizontal curve about which the five-brane wraps.

Our main interest in this paper is the question: \emph{How do the bundle
moduli
change during a gauge changing phase transition?} As such transition is
naturally associated
to reducible spectral covers \cite{FMW, BJPS, OPP} and appears to be
hard to study in general, we will
adopt here the situation (first studied in \cite{BJPS}) of having a vector
bundle $E=V\oplus \pi^*M$. Here
$M$ is a good vector bundle on the base of $X$ whose second Chern
class counts the five-branes dissolved in the transition. The
associated spectral
cover is the union of the spectral cover of $V$ and
the zero section $m\sigma$ (which carries the rank $m$ vector bundle
$M$). If one
considers the structure group of the involved bundle one encounters the
following situation: one
starts with a stable $SU(n)$ vector bundle $V$ on $X$ whose structure
group changes
during the transition to $SU(n)\times SU(m)$ as discussed in \cite{OPP}.

It is known that the moduli of $E$ decompose into four classes: the
moduli of $V$, $\pi^*M$ and
the moduli which `measure' the deviation of $E$ from being a direct
sum. The main problem which
occurs is that only the difference of the latter moduli can be obtained
by an index computation
on $X$ \cite{BJPS} which will be reviewed in section 
4.4.  Now, two
observations will help us to
obtain information about the total number of the moduli. First, the
mentioned index which one can
evaluate on $X$ is proportional to the net-generation number (which is
one half of $c_3(V)$); second,
chiral matter is localized along the intersection curve $S$ of $C_V$
and $\sigma$ (first pointed out in
\cite{FMW} and later used in \cite{C} and \cite{DI}). These observations and the fact
that $E$ can be obtained by
a Fourier-Mukai transformation lead to the idea of reducing the
index computation on $X$ to an
index computation on the intersection curve $S$ as we will explain in
section 
4.5. This reduction
will help to apply a vanishing argument and allows to obtain information
about the total number of moduli.

Finally in section 5, we will apply a similar argument to the
computation of the
number of chiral matter multiplets in heterotic compactifications on
ellitically fibered
Calabi-Yau threefolds $X$.
As a result, we find an alternative derivation of the net-generation
number as originally performed in \cite{DI} and show that if a vanishing
argument applies, one
can obtain the precise number of chiral matter multiplets. We find that
chiral matter associated to
$H^1(X,V)$ vanishes in heterotic string compactifications on
elliptically fibered $X$
with vector bundles constructed in the spectral cover approach and
matter
localized along a curve $S$ of arithmetic genus $g(S)>1$. This implies
the vanishing of
the corresponding Yukawa couplings.

Let us summarize the organization of this paper. In section 2, we
review the
spectral cover construction of vector bundles. In section 3, we prove
that the
Fourier-Mukai transformation preserves absolute stability. In section
4, we first review the necessary
facts about the five-brane instanton transition. Then we work out a
formula for the number of moduli which occur in a (vertical) five-brane
instanton transition.  We show how this moduli can be explicitly
computed by applying a vanishing argument. In section 5, we reconsider
the localization of chiral matter. We apply a similar vanishing argument
to compute the number of chiral matter multiplets.
The appendices contain all necessary calculations and proofs required
for the sub-sections.

\section{Review of Vector Bundle Construction}
We begin by recalling the construction of  vector bundles on elliptic
Calabi-Yau
threefolds following the original construction given in
\cite{FMW} to which we refer for more details.

\subsection{Spectral Cover Construction}\ 

\medskip
One starts with a Calabi-Yau threefold $X$ which is elliptically
fibered over a complex
two-dimensional base $B$ and denotes by $\pi$ the projection of $X$
onto $B$. In addition
one requires that $X$ has a section $\sigma$. The construction of
$SU(n)$ vector bundles $V$ (fiberwise semistable and with $c_1(V)=0$)
on $X$
using the spectral cover construction proceeds in two steps. First one
describes bundles on the elliptic fiber and then uses
global data in the base to ``glue'' them together to a bundle $V$ on $X$.

More precisely, one starts on an elliptic fiber $F$ given in the
Weierstrass representation
with distinguished reference point $p$. On the fiber, $V$ decomposes as
a direct sum of degree zero line bundles, each associated with a unique
point on $F$. The condition that $V$ is an
$SU(n)$ bundle means that the product of the line bundles is trivial or
equivalently
that the points sum up to zero in the group law on $F$. For this
$n$-tuple
of points  exists a meromorphic function vanishing to first order at
the points and
having a pole only at $p$.

When the reference point is globalized by the section $\sigma$ the
variation of the $n$
points in the fiber leads to a hypersurface $C$ embedded in $X$ which
is a ramified
$n$-fold cover -the spectral cover- of the base given by
\begin{equation}
s=a_0+a_2x+a_3y+...+a_nx^{n/2}=0
\label{spec}
\end{equation}
here
$a_r\in\Gamma(B,{\M}\otimes K_B^r)$, $a_0$ is a section of $\M$ and $x,y$ sections of $K_B^{-2}$ resp. $K_B^{-3}$ in the
Weierstrass model of $X$ \cite{FMW}. Note the last term \ref{spec} 
for $n$-odd is $a_nx^{(n-3)/2}$. The pole order
condition leads to
$s$ being a section of $\Oc(\sigma)^n$ which can in the process of
globalization still
be twisted by a line bundle ${\M}$ over $B$ of $c_1({\M})=\eta$. Thus $s$ can be actually
a section of ${\Oc}(\sigma)^n\otimes {\M}$  and the cohomology
class of $C$ in $X$ is
$$
C=n\sigma+\pi^*\eta
$$
So far we have recalled how to construct a spectral cover $C$ by
starting with a vector bundle
$V$ over $X$. The basic idea of the spectral cover construction is now
to recover $V$ from $C$!
Therefore one starts with a suitable line bundle ${\Rcal}$\footnote{If $C$
is not irreducible,  ${\Rcal}$ may be only a sheaf of rank one with no
concentrated subsheaves} on the $n$-fold cover $p\colon X\times_B C
\rightarrow X$ and $V$ will be induced as $V=p_*{\Rcal}$. If one takes
for ${\Rcal}$ the
Poincar\'e sheaf ${\Pc}$ on $X\times_B X$ (suitably modified in away
that will made precise in the following section) and takes into account
that the twist by a line
bundle $L$ over $C$ leaves the fiberwise isomorphism class unchanged,
one obtains
\begin{equation}V=p_*(p^*_CL\otimes {\Pc})
\label{fms}
\end{equation}
where $p$ and $p_C$ are the projections of the first and second factor
of $X\times_B C$.
The condition $c_1(V)=0$ translates to a fixing of $\pi_*c_1(L)$ in
$H^{1,1}(C)$ up to a class
in $ker \pi_*\colon H^{1,1}(C)\rightarrow H^{1,1}(B)$.

\subsection{Fourier-Mukai Transformation}\ 

\medskip
The structure of $V$ which occurs in \ref{fms} makes transparent that $V$
can be considered
more generally as  Fourier-Mukai transformation of the pair $(C,L)$.
For the description
of the Fourier-Mukai transform it is appropriate instead of working on
$X\times_B C$ to work on
$X\times_B {\tilde X}$ where $\tilde X$ is the compactified relative
Jacobian of $X$. $\tilde X$ parameterizes torsion-free rank 1 and
degree zero sheaves of the fibers of $X\to B$  and it is actually
isomorphic with $X$  (see \cite{bart} or \cite{RPo}) so that we will identify
$\tilde X$ with $X$.

We have a diagram:
$$
\xymatrix{ X\times_B{X} \ar[r]^{p_2}\ar[d]^{p_1} & X \ar[d]^{\pi_2} \\
X \ar[r]^{\pi_1} & B}
$$
and the \emph{Poincar\'e} sheaf
$$
{\Pc}={\Oc}(\Delta)\otimes {\Oc}(- p_1^*\sigma)\otimes {\Oc}(-p_2^*\sigma)\otimes q^*K_B^{-1}
$$
normalized to make ${\Pc}$ trivial along $\sigma \times \tilde{X}$and
$X\times \sigma$. Here
$\sigma$ is the fixed section, 
 $q=\pi_1\circ p_1=\pi_2\circ p_2$ and ${\Oc}(\Delta)$ is the dual of the ideal sheaf of the diagonal, which is
torsion-free of rank 1.

The Fourier-Mukai transform and the inverse Fourier-Mukai transform are
defined as functors of the derived categories $D(X)$ of complexes
of coherent sheaves on $X$ bounded from above. We have
\begin{align*}
\varPhi&\colon D^{-}(X)\rightarrow D^{-}(X)\,; & 
\varPhi(\G)&=Rp_{1*}(p^*_2(\G)\otimes \Pc)\,, \\
\hat{\varPhi}&\colon
 D^{-}(X)\rightarrow D^{-}(X)\,; 
& \hat{\varPhi}(\G)=Rp_{2*}(p^*_1(\G \otimes \hat{\Pc})
\end{align*}
where
$$
\hat{\Pc}={\Pc}^*\otimes q^*K_B^{-1}\,.
$$
We can also define the Fourier-Mukai functors ${{\varPhi} }^i$ and
${\hat {\varPhi} }^i$,
$i=0,1$ in terms of single sheaves by taking ${\varPhi}  ^i({\F})$
and $\hat {\varPhi}  ^i({\F})$
as the $i$-th cohomology sheaves of the complexes ${\varPhi}  ({\F})$ and $\hat
{\varPhi}  ({\F})$, we have
\begin{align}
{\varPhi}  ^i({\F})&=R^ip_{1*}(p_2^*({\F})\otimes {\Pc})\,,
\\
\hat {\varPhi}  ^i({\F})&=R^ip_{1*}(p_2^*({\F})\otimes
\hat{\Pc})\,.
\end{align}

\subsubsection*{WIT$_i$ Sheaves}\ 

 \smallskip

We can now talk about WIT$_i$ sheaves: they are those
sheaves ${\F}$ for which ${\varPhi}  ^j({\F})=0$ for $j\neq
i$, and we
have the same notion for the inverse Fourier-Mukai transform.

Note also that the Fourier-Mukai transform and the inverse Fourier
Mukai transform
are only inverse functors up to a shift  (see for instance \cite{RPo}, Lemma
2.6):
$$
{\varPhi}  (\hat{{\varPhi}  }({\G}))= {\G}[-1]\,,\quad
 \hat{{\varPhi}  }({{\varPhi}  }({\F}))= {\F}[-1]\,.
$$
The -1 shift implies that if we have a single sheaf ${\F}$, then
$\hat{{\varPhi}  }({{\varPhi}  }({\F}))= {\F}[-1]$ is a
complex with only one cohomology sheaf,
which is ${\F}$, but located at ``degree1''. Now if
${\varPhi}  ^0({\F})=0$ (${\F}$ is WIT$_1$) then the unique
Fourier-Mukai transform ${\varPhi}  ^1({\F})$ is
WIT$_0$ for the inverse Fourier-Mukai transform such that
$\hat{{\varPhi}
}^0({\varPhi}  ^1({\F}))={\F}$. In the
same way, if ${\varPhi}  ^1({\F})=0$ (${\F}$ is WIT$_0$) then
${\varPhi}  ^0({\F})$ is WIT$_1$ for
the inverse Fourier-Mukai and $\hat{{\varPhi}  }^1({\varPhi}
^0({\F}))={\F}$. Let us now return to the spectral cover
construction.

Given a relatively semistable vector bundle $V$ of rank $n$ on $X$,
\footnote{More
generally we can take $V$ flat over $B$ and torsion-free semistable of
degree 0 on
fibers.} then $V$ is WIT$_1$ and the unique Fourier-Mukai transform
${\varPhi}  ^1(V)$ is supported on a surface $i\colon C\to X$ inside
$X$ and
its restriction to $C$ is a  pure dimension 1 and rank 1 sheaf $L$ on
$C$.  That is,
${\varPhi}  ^1(V)=i_*L$.

The surface $C$  projects onto the base $B$ as a $n:1$ cover, the \emph{spectral cover} of $V$.  Due to the invertibility of the
Fourier-Mukai transform,
$i_*L$ is WIT$_0$ for the inverse Fourier-Mukai so that we can recover
the bundle
$V$ in terms of the spectral data as
$$
V=\hat {\varPhi}  ^0(i_*L)=p_{1*}(p_2^*(i_*L)\otimes \hat{\Pc})\,.
$$
We can follow the inverse road: take a surface $i\colon C\hookrightarrow X$ inside
$X$ flat over $B$, and a  pure dimension 1 and rank 1 sheaf $L$ on $C$
(for instance a line bundle). Then $i_*L$ as a sheaf on $X$ is WIT$_0$
for the inverse Fourier-Mukai transform. Its inverse $V=\hat {\varPhi}^0(i_*L)$ is a
sheaf on $X$ relatively torsion-free semistable and of degree 0
\cite{FMW, OUR, RPo}.

The topological invariants of ${\varPhi} (\G)$ and $\hat
{\varPhi}  (\G)$ for an
arbitrary object $\G$ of the derived category has been computed
explicitly in terms of those of $\G$ in \cite{OUR}.

\section{Comments on Stability}

For a line bundle $L$ on an irreducible spectral cover $C$ the
Fourier-Mukai transform of $i_*L$
is a stable vector bundle $V$ on $X$. But sometimes we have
to deal with vector bundles (or more generally torsion-free sheaves)
whose spectral cover $C$ is not irreducible such that $L$ is no
longer a line bundle on $C$ but rather a torsion-free rank one sheaf.
It follows $L$ is not automatically stable. We will show that (for
$V$ semistable of degree zero on fibers) the stability of $L$ (as a
torsion-free rank one sheaf on the reducible spectral cover) is
equivalent to the stability of $V$ as a torsion-free sheaf on $X$.
As a result, Simpson Jacobians of stable torsion-free rank one sheaves
on
spectral covers are isomorphic to open subsets of moduli spaces of
stable sheaves on $X$.

\subsection{Review of the Elliptic Surface Case}\ 

\medskip
It is known that for elliptic surfaces the relative Fourier Mukai
transform preserves not only fiberwise stability (see \cite{bart} for the
case of positive degree on fibers and \cite{RPo} for the case of degree zero
on fibers)
but also absolute stability in a certain sense. By fiberwise or
relative stability (or
semistability) we understand stability (or semistability) on fibers.
A sheaf ${\F}$ on a fibration $X\to B$ is said to be fiberwise
or relatively stable (or semistable) if it is flat over $B$ and the
restriction of ${\F}$ to every fiber of $X\to B$ is stable (or
semistable) in the ordinary sense. Further we need a relative
polarization to
speak about relative stability. Such polarization is given by a divisor
on $X$ that
meets every fiber in a polarization of the fiber.

Relative stability is a very important concept, however, when we have
our
elliptic fibration $X\to B$ we need to consider ``absolute stability''
as well. Here we refer to stability on $X$ with respect to a
certain polarization; we then somehow forget the fibered structure
and consider $X$ just as a manifold. The reason we need
absolute stability is that we want to consider moduli spaces of
stable sheaves on $X$.

There is still one more thing to keep in mind. Stability (or
semistability) used to be defined in terms of the slope
($\mu$-stability) or the Hilbert polynomial (Gieseker stability) but
only for torsion-free sheaves. That excluded sheaves concentrated on
closed subvarieties or even sheaves defined on reducible singular
varieties. This problem was circumvented by Simpson \cite{Simp} who defined
both $\mu$-stability and Gieseker stability (along with the
corresponding semistability notions) for ``pure'' sheaves on arbitrary
projective varieties. For Simpson, a pure sheaf of dimension $i$ is a
sheaf ${\F}$ whose support has dimension $i$ and that has no
subsheaves concentrated on smaller dimension. 
This gives the more natural generalization of the notion of torsion-free.

Now recall that the Euler characteristic of a vector bundle (or more
general, coherent sheaf)
${\F}$ is given by $\chi({\F})=\sum(-1)^i\dim H^i(X,{\F})$
and for a fixed ample line
bundle ${\Oc}(1)$ on $X$ we have the so called Hilbert polynomial
$P({\F},m)$ given by ${m\rightarrow \chi({\F}\otimes {\Oc}m)}$.
The Hilbert polynomial can be written as
$$
P({\F},m)=\sum_{i=0}^3 \alpha_i({\F}) \frac{m^i}{i!}
$$
with integral coefficients $\alpha_i({\F})$ which are listed in the
appendix.
Once the Hilbert polynomial (and then the slope) is defined, Simpson
definitions are very similar to the ordinary ones.

When the support $Y$ of a pure sheaf ${\F}$ is irreducible and
reduced so that the restriction ${\F}_{\vert Y}$ is torsion-free on
$Y$,
the stability of ${\F}$ as a sheaf on $X$ in the sense of Simpson
is equivalent to the stability of ${\F}_{\vert Y}$ as a torsion-free
sheaf on $Y$ in the ordinary sense. But when $Y$ is reducible, we have
to consider Simpson stability as the unique reasonable notion. This is
not an uncommon situation, for instance, if we take an elliptic
fibration $X\to B$ and  a relatively semistable sheaf ${\F}$ on
$X$ of degree zero on fibers, then the Fourier-Mukai transform
${\varPhi}^1({\F})$ is concentrated on the spectral cover $C$ that
in many cases is reducible. We can thus only consider the possible
stability of ${\varPhi}^1({\F})$ in the sense of Simpson.

By this reason, in what sequel stability will always mean
$\mu$-stability in the sense of Simpson.

As a warm up, we start the discussion about preservation of absolute
stability in the
case of an elliptic surface $X\to B$. Now, given a Cartier divisor
$C\hookrightarrow\widehat X$ flat of degree $n$
that we polarize with the intersection $F_C$ of $C$ with the fiber
$F$ of $\pi$, we have the following result:  for every  $a>0$, there
exists $b_0\ge 0$ depending only on the topological
invariants of $C$, such that for every
$b\ge b_0$ and every  sheaf $L$ on $C$ of pure dimension one,
rank one, degree $r$ and semistable with respect to $\mu_C$, the
unique Fourier-Mukai transform ${\varPhi}^0({L})$ is semistable
on $X$ with respect to the
polarization $a\sigma+b F$. Moreover, if $L$ is stable on
$C$, then ${\varPhi} ^0(L)$ is stable as well on $X$.

A certain converse is also true: let us fix a Mukai vector
$(n,\Delta,s)$ with $\Delta\cdot F=0$. For every $a>0$, there exists
$b_0$ such that for every $b\ge b_0$ and every sheaf $V$ on
$X$ with  Chern character $(n,\Delta,s)$ and semistable with respect to
the polarization $a \sigma+b F$, the restriction of $V$ to the generic
fiber is
semistable. In particular, $V$ is  WIT$_{1}$ so that it has a unique
relative Fourier-Mukai transform $\hat{{\varPhi}}^1(V)$. If we
assume that the restriction of $V$ to every fiber is semistable then
$\hat{{\varPhi}}^1(V)$ is of pure dimension one, rank one, degree $r$
and semistable on the spectral cover $C(V)$. In other words
the spectral cover $C(V)$ does not contain fibers. If $V$ is stable on
$X$, $\hat{{\varPhi}}^1(V)$ is stable on $C(V)$ as well (see \cite{RPo}).
These properties mean that we can construct non-empty open subsets of
components of the moduli space of stable sheaves on $X$ in terms of the
compactified Simpson Jacobians of the spectral covers \cite{RPo}.\footnote{For an
integral
variety, the Jacobian parameterizes line bundles of a fixed degree. In
that case, line bundles are automatically stable, regardless of the
polarization. In the reducible case this is no longer true,
we need to fix a polarization and we can have line bundles that are
unstable. In this situation instead of parameterizing line bundles, we
parameterize pure dimension $n$ (the dimension of the space) rank one
and fixed degree sheaves that are stable in the sense of Simpson. The
corresponding moduli space is the compactified Simpson Jacobian. For an
integral variety, this is a compactification of the ordinary Jacobian
because it contains not only all line bundles, but also the
torsion-free rank one sheaves.}

\subsection{Generalization to Calabi-Yau Threefolds}\ 

\medskip
We want to study now a similar question for elliptic
Calabi Yau threefolds $\pi\colon X\to B$. Since
we require that $\pi$ has a section (in addition to the smoothness
of $B$ and $X$), the base surface $B$ has to be of a particular kind,
namely $B$ has to be Del Pezzo, Hirzebruch, Enriques or a blow-up of a
Hirzebruch (see \cite{DLO} or \cite{mv}).

To our knowledge the problem of preservation of
absolute stability for elliptic Calabi-Yau threefolds has not been
considered
in the literature so far. We polarize $X$ with $\tilde H=a\sigma +b H_B$,
where
$H_B=\pi^*(\bar H_B)$ and $\bar H_B$ is a polarization of $B$ to be chosen later.
In the following we will assume that there is a decomposition
\begin{equation}
H^{2i}(X)=\sigma p^* H^{2i-2}(B)\oplus p^* H^{2i}(B)
\label{decomposition}
\end{equation}

Let us consider a torsion-free sheaf $V$ on $X$ of rank $n$ and degree
zero on fibers and write its Chern characters as
$ch(V)=(n,\tilde S,\sigma\eta+aF,s)$ with $\eta, \tilde S\in
p_2^*H^2(B)$
according with \cite{OUR} and \ref{decomposition}.
 Assume that $V$ is WIT$_1$. This happens, for
instance, when $V$ is relatively semistable (i.e., flat over $B$ and
semistable on fibers). The Hilbert polynomial of the unique
Fourier-Mukai
transform $\hat{{\varPhi}  }(V)$ of $V$ is given by $P(\hat{{\varPhi}
  }(V),m)$
with $\alpha_i(\hat{{\varPhi}  }(V))$ given in appendix A.
Simpson slopes of $V$ and ${\varPhi}  (V)$ can then be determined and
are given by
\begin{equation}
\mu(V)=\frac{\tilde S\cdot \tilde H^2}{n\cdot\tilde H^3}\,,
\quad
\text{and}\quad\mu({\varPhi}  (V))=\frac{\alpha_1(\hat{{\varPhi}
}(V))}{\alpha_2(\hat{{\varPhi}  }(V))}
\label{slo}
\end{equation}
We make now a further assumption, that the support $C$ of
${\varPhi}(V)$
is flat over $B$, that is, that it does not contain fibers of $\pi$. It
follows that
the support of every subsheaf ${\F}$ of ${\varPhi}  (V)$ is
contained in $C$ and has no fibers as well, so that it is WIT$_0$ with
respect to the inverse Fourier-Mukai transform and its transform is a
WIT$_1$
subsheaf $V'$ of $V$. Moreover, $V'$ has degree zero on fibers again by
\cite{OUR} (2.33)
so that \ref{slo} is still true for $V'$ (writing primes for the
correspondent invariants).

For every $a$ there exists $b>0$, depending only on the topological
invariants of $V$ such that $V$ is stable (resp. semistable) with
respect to $\tilde H=a\sigma+bH_B$ if and only if ${\varPhi}  (V)$ is
stable (resp.
semistable) as well. Assume that $V$ is stable and that ${\varPhi}
(V)$ is destabilized
by a subsheaf ${\F}$. Then, as we said before, ${\F}={\varPhi}(V')$ for
certain subsheaf $V'$ of $V$ of degree zero on fibers and we have
$$
\frac{\alpha_1(\hat{{\varPhi}}(V))}{\alpha_2(\hat{{\varPhi}  }(V))}\leq
\frac{\alpha_1(\hat{{\varPhi}  }(V'))}{\alpha_2(\hat{{\varPhi}
}(V'))}
$$
with the $\alpha_i$'s given in appendix A. 

If we write this condition as a polynomial on $b$, we have
\begin{multline*}
n'(\sigma H_B^2)(\sigma \tilde S H_B+ \frac12 n c_1\sigma
H_B)b^3+\text{lower terms}
\\
\leq
n(\sigma H_B^2)(\sigma \tilde S' H_B+ \frac12 n' c_1\sigma
H_B)b^3+\text{lower terms.}
\end{multline*}
Since the family of subsheaves of $V$ is bounded,
there is a finite number of possibilities for the Hilbert polynomial of
$V'$. Then, the value of $b$ one has to chose depends only on $a$ and
on the topological invariants of $V$. For $b\gg 0$ the destabilizing
condition
is equivalent to
$$
n'(\sigma \tilde S H_B)\leq n(\sigma \tilde S' H_B)
$$
On the other hand the stability of $V$ gives
${n (\tilde S'\cdot \tilde H^2) < n'(\tilde S\cdot \tilde
H^2)}$
and then
$$
n' (\sigma c_1 \tilde S) < n (\sigma c_1 \tilde S')
$$
which is a contradiction. Both the converse and the corresponding
semistability statements are proven analogously.

For elliptically fibered Calabi-Yau threefolds $X$
we proceed as in the elliptic surface case and basically
use the above result to prove that there are open subsets of
components of the moduli space of stable sheaves on $X$ that are
isomorphic to compactified Simpson Jacobians of a universal spectral
cover.

A last comment: we have shown the preservation of absolute stability
under Fourier Mukai transformation for sheaves of degree zero on fibers.
In this paper, we shall focus only on fiber bundles with vanishing first
Chern class $c_1(V)=0$, because they are the ones relevant for the
problems we are considering. Their Fourier-Mukai transforms
${\varPhi}^1(V)$ have non-vanishing $c_1$ ($c_1({\varPhi}^1(V))$
which represents the spectral cover of $V$; because of this,  the
problem of
preservation of stability is better studied in the more symmetric
situation of degree zero on fibers, a property shared by $V$ and
${\varPhi}^1(V)$.

\section{Five-Brane Instanton Transition}

\subsection{Anomaly Cancellation}\ 

\medskip
Heterotic string compactifications on elliptic Calabi-Yau threefolds
require a number of five-branes in order to cancel the anomaly
\cite{FMW}.
These five-branes wrapping holomorphic curves in $X$ whose cohomology
class is determined by the heterotic anomaly cancellation condition
\begin{equation} [W]=c_2(TX)-c_2(V_1)-c_2(V_2)
\label{hac}
\end{equation}
where $[W]$ is the cohomology class of the wrapped curves, $c_2(V_i)$
are
the second Chern classes of the vector bundles on $X$, $c_2(TX)$ is
the second Chern class of the tangent bundle and given by \cite{FMW}
$$
c_2(TX)=c_2+11c_1^2+12\sigma c_1\,.
$$
We use the notation $c_i=\pi^*c_i(B)$ and $\sigma$ (satisfying
$\sigma^2=-c_1\sigma)$
the class of a section of $\pi$. 
Due to our assumption \ref{decomposition},
$[W]$ may be
decomposed as
$$
[W]=\sigma C_1+C_2
$$
where $C_1$ maps to a divisor in $B_2$ to be embedded in $X$ via
$\sigma$ and $C_2=h[F]$ describes the
five-branes wrapping the elliptic fiber of $X$. Following \cite{DIA} we
refer to five-branes which wrap curves in the base as horizontal
five-branes
and branes wrapping the fiber as vertical ones. We will also mention
that five-branes
could wrap skew curves, i.e. curves which have both fiber and base
components.

\subsection{The Transition}\ 

We will be interested now in the situation when heterotic five-branes
which wrap the
elliptic fiber dissolve into gauge instantons resulting in a new
heterotic vector bundle $E$.
Let us follow how such transition might proceed thereby recalling
results partly obtained in \cite{BJPS} and \cite{OPP}. To simplify our discussion
we will
assume (until otherwise stated) that $V_2$ in \ref{hac} is trivial. Under
heterotic/F-theory duality this corresponds to an unbroken
$G=G_1\times E_8$ gauge group where $G_1$ is determined by the
commutator of the structure group of $V_1$ in $E_8$. Further we assume
that $[W]=hF$, that is, we only consider `vertical five-branes'.

On the level of anomaly cancellation one expects that a five-brane
instanton transition
causes a change in the second Chern-class of the vector bundle
$$
c_2(TX)-(c_2(V)+kF)=[\tilde W]
$$
assuming here $k< h$ so `absorbing' part of the five-brane class into
the vector bundle.

Now we can think of a five-brane at the transition point (a particular
point on the Coulomb branch)
as a pointlike instanton concentrated in codimension two in $X$ which
will be the elliptic fiber in our case. That is, a five-brane can be
considered as singular gauge field configuration such that the
curvature is zero
everywhere except on a fiber where it has a singularity. In terms of
Hermitian-Yang-Mills connections we can think of a connection on a
vector bundle which is smooth except along a curve in the class $C_2$
where it has a delta function behavior. Mathematically such a
configuration is described by a singular torsion free sheaf.
If the singular sheaf can be smoothed out to a vector bundle a
five-brane instanton transition can occur.

If we recall that $c_2(V)$ is in $H^4(X)$ and as we are concerned with
an elliptic fibration $\pi\colon X\rightarrow B$ we have a decomposition
${H^4(X)=H^2(B)\sigma\oplus H^4(B)}$ with $\sigma$ being the section.
For $c_2(V)$ one has ${c_2(V)=\pi^*(\eta)\sigma+\pi^*(\omega)}$ with
$\eta, \omega\in H^2(B)$ resp. $H^4(B)$. Therefore it is expected
\cite{BJPS} that the
singular configuration can be smoothed out
to a new bundle with $\int_B c_2(E)=\int_B c_2(V)+k$ assuming that $kF$
can be represented by $k$ separated fibers projecting to $k$ distinct
points on $B$. This suggests that we are looking for a vector bundle
$M$ (or sheaf) on $B$ with $c_2(M)=k$.

Thus the actual transition proceeds in two steps \cite{OPP}. First, one
describes a
singular torsion free sheaf $\tilde M$ on $B$. Second, one shows that
it can be smoothed out to a stable bundle $M$ on $B$ which pulls back
to a stable vector bundle over $X$ with $c_2(\pi^*M)=kF$.

To summarize: after the transition a non-trivial gauge bundle $M$
of rank $m$ has developed on the zero section. The new bundle
$E=V\oplus \pi^*M$ is smooth and reducible of rank $n+m$.
The second Chern class of $M$ counts the number of five-branes which
have been dissolved in the transition.  As the zero section $\sigma$ is
isomorphic to the base, one can think of $M$ as being a vector bundle
on $B$.

The spectral cover of $E=V\oplus \pi^*M$ can be easily described in
terms of the spectral covers of $V$ and $\pi^*M$. One notices that $E$
is still WIT$_1$ and that its unique Fourier-Mukai transform is the
direct sum ${\varPhi}  ^1(E)={\varPhi}  ^1(V)\oplus {\varPhi}
^1(\pi^*M)$ of
the Fourier-Mukai transforms of $V$ and $\pi^*M$.
We know (\cite{RPo}) that the spectral cover is closed defined by the
Fitting ideal of the Fourier-Mukai transform. Since the Fitting ideal
which describes the spectral cover is multiplicative over direct sums
(see \cite{RPo}), the spectral cover $C_E$ is the union of the spectral covers
$C_V$ and $C_{\pi^*M}$, that is
$$
C_E=C_V+C_{\pi^*M}
$$
as numerical classes. If we proceed as in \cite{RPo}\footnote{The
Poincar\'e sheaf considered in \cite{RPo} is the dual of the Poincar\'e
sheaf considered both here and in \cite{OUR}.}  one obtains
\begin{equation}
{\varPhi}  ^1(\pi^*M)=\pi^*M\otimes {\varPhi}  ^1({\Oc}_X)=\pi^*M\otimes\pi^*K_B\otimes
{\Oc}_\sigma=\sigma_*(G)\,,\quad G=M(K_B)
\label{gsheaf}
\end{equation}
Then ${\varPhi}^1(\pi^*M)=\sigma_*(G)$ is concentrated on $\sigma$,
but due to the multiplicativity of the Fitting ideal, the spectral cover
of $\pi^*M$ is not $\sigma$ but rather $C_{\pi^*M}=m\sigma$ and the
restriction  $L_M=\sigma_*(G)_{\vert  m\sigma}$ of the Fourier-Mukai
transform to the spectral cover is a pure dimension one rank one sheaf
 on the reducible surface
$m\sigma$ (which is not a line bundle). We then have
$$
C_E=C_V+m\sigma
$$
as we expected and
$$
{\varPhi}  ^1(\pi^*M)=\sigma_*(G)=h_*(L_M)
$$
where $h\colon m\sigma\hookrightarrow X$ is the  immersion of $m\sigma$ into $X$.

\subsubsection*{Matter Curve $S$}\ 

\smallskip

It is known that chiral matter ($c_3(V)/2\neq 0$) is localized along the intersection curve $S$ of $\sigma$ and $C$ \cite{FMW, DI}. 

In the following we will assume that
\begin{itemize}
\item $S=\sigma\cdot C$ is irreducible
\item $g(S)>1$.
\end{itemize}

\subsection{F-Theory Perspective}\ 

\medskip
Let us recall how the five-brane instanton transition is viewed
from the perspective of F-theory!

Recall that F-theory is defined as type-IIB super string theory with
varying coupling constant \cite{V}.
A consistent F-theory compactification on an Calabi-Yau fourfold
$Y$\footnote{here assumed to be elliptically
fibered over a three dimensional base $B$ which is a ${\Ps}^1$ bundle
over the same $B_2$ as considered on the heterotic string side}
requires a number $\chi(Y)/24$ of three-branes  filling the
transverse space time $\R^{3,1}$ \cite{SVW}.
This number agrees with the number of five-branes required for
consistent F-theory compactification
giving a non-trivial test for the expected heterotic/F-theory duality
\cite{FMW, AC} and the adiabatic argument.

Now by duality one expects that if a five-brane disappears on the
heterotic side, a three-brane should disappear on the F-theory side.
More precisely, a three-brane `dissolves' into a finite size instanton,
i.e.
a background gauge bundle $\tilde M$ on the corresponding component
on the seven-brane is turned on \cite{BJPS}. The instanton number of this
bundle
counts thereby the number of dissolved three-branes. In \cite{BJPS} it was
then
suggested that the two bundles $M$ and $\tilde M$ should actually be
identified.

Such a transition leads to a modification of the anomaly cancellation
condition
$$
\frac{\chi(Y)}{24}=n_3+\sum_{j}k_j
$$
here $n_3$ is the number of three-branes and $k_j=\int_{D_j}c_2(M_j)$
and $j$ labels the respective component of the seven-brane partially
wrapped over the discriminant locus\footnote{\cite{BJPS} argues that a
three-brane can only dissolve on a multiple seven-brane} in the three
dimensional base of $Y$.

Due to the presence of a non-trivial instanton bundle on the
seven-brane part of the gauge group will be broken. Otherwise, the gauge
group would be given by a degeneration of A-D-E type of the elliptic
fiber
over the compact part of the seven-brane. The gauge group which is left
over after
the breaking by $\tilde M$ should correspond on the
heterotic side to the commutator of $E$ in $E_8$. Further one expects
that in the transition extra chiral matter occurs (if the original
bundle $V$ had non zero $c_3(V)$) \cite{BJPS} which is on the heterotic side
related to the moduli we are aiming to evaluate.

\subsection{Computation of Moduli}\ 

\medskip
To begin let us recall some known facts about the moduli of $E=V\oplus
\pi^*M$.
The number of moduli is given by the dimension of the deformations
space $H^1(\End (E))$.
This space can be decomposed into four parts  (as already noticed in
\cite{BJPS})
\begin{align*}
H^1(\End (E))= & H^1(\End (V))\oplus
H^1(\End(\pi^*M))\oplus  \\
 &   H^1(Hom(V,\pi^*M))
\oplus H^1(\Hom(\pi^*M,V))
\end{align*}
where the first two summands correspond to deformations of $E$ that
preserve the direct sum and deform $V$ and $\pi^*M$ individual. The
last two elements give the
deformations of $E$ that deform away from the direct sum.

\subsubsection*{Moduli of $V$}\ 

\smallskip

The number of moduli of $V$ can be determined in two ways, depending
on whether one works with $V$ directly or with its spectral cover data
$(C,L)$ from which it is obtained. In the direct approach one is restricted to
so called $\tau$-\emph{invariant} bundles and therefore to a rather special point
in the moduli space, whereas the second approach is not restricted to such a
point. However, after a brief review of the first approach which was originally
introduced in \cite{FMW} making concrete earlier observations in \cite{Witnew}, we will
explain how the $\tau$-\emph{invariance} is translated to the $(C,L)$ data. The issue
of $\tau$-\emph{invariance} has been also addressed in \cite{DoOv}.

The first approach starts with the index of the $\bar\partial$ operator
with values in $\End(V)$ which is the $\operatorname{index}(\bar\partial)=\sum_{i=0}^{3}(-1)^i \dim H^i(X,\End(V))$.
As this index vanishes by Serre duality on the Calabi-Yau threefold,
one has to introduce
a further twist to get a non-trivial index problem. This is usually
given if the Calabi-Yau space
admits a discrete symmetry group \cite{Witnew}. In case of elliptically
fibered Calabi-Yau manifolds
one has such a group $G$ given by the involution $\tau$ coming from the
``sign flip'' in the elliptic
fibers. Now one assumes that this symmetry can be lifted to an action
on the bundle at least at some
point in the moduli space \cite{FMW}. In particular the action of $\tau$
lifts to an action on the adjoint bundle $ad(V)$ which are the
traceless endomorphisms of $\End(V)$. It follows that the index of the
$\bar\partial$ operator generalizes to a character valued index where
for each $g\in G$ one defines
$\op{index}(g)=\sum_{i=0}^3(-1)^{i+1} \op{Tr}_{H^i(X,ad(V))}g$ where
$\op{Tr}_{H^i(X,ad(V))}$
refers to a trace in the vector space $H^i(X,ad(V))$. The particular
form of this index for elliptic Calabi-Yau threefolds has been
determined in \cite{FMW} (with $g=1+\frac{\tau}2$) one finds $\op{index}(g)=
\sum_{i=0}^3(-1)^{i+1} \dim H^i(X, ad(V))_e$ where the subscript
``e'' indicates the projection
onto the even subspace of $H^i(X, ad(V))$. One can compute this index
using a fixed point theorem
as shown in \cite{FMW}.

Now the second approach makes intuitively clear where
the moduli of $V$ are coming from, namely, the number of parameters
specifying the spectral cover $C$ and by the dimension of the space of
holomorphic
line bundles $L$ on $C$. The first number is given by the dimension of
the linear system
$|C|=|n\sigma+\eta|$. The second number is given by the dimension of
the Picard group $Pic(C)=H^1(C,{\Oc}^*_C)$ of $C$.
One thus expects the moduli of $V$ to be given by \cite{BDO}
$$
h^1(X, \End(V))=\dim |C|+\dim Pic(C)
$$
which can be explicitly evaluated making the assumption that $C$ is an
irreducible, effective, positive divisor in $X$.

If one computes the endomorphisms of $V$ using the character valued
index
one assumes that $V$ is invariant under the involution of the elliptic
fiber, i.e.$V=V^\tau$.
On the other hand the number of moduli derived from the pair $(C,L)$
requires no such restriction.
Thus the question occurs: \emph{how is the condition $V=V^\tau$
translated to the spectral data $(C, L)$?}

To see this  translation let us study the meaning of the condition
$V^\tau=V$ with respect to the relative
Fourier Mukai transformation $\varPhi$, that is to
$$
\varPhi\colon D(X)\to D(X)\,,\quad F\mapsto
\varPhi(F)=\pi_{2,*}(\pi_1^*(F)\otimes \Pc)\,,
$$
where $\Pc$ is the relative Poincar\'e sheaf on $X\times_B X$. As already
mentioned, $\varPhi$ is an equivalence of categories whose inverse functor is
the Fourier Mukai transform $\hat \varPhi$ with respect to $\Pc^*\otimes
q^*K_B^{-1}$, where $q\colon X\times_B X\to B$ is the natural
projection and $K_B$ is the canonical sheaf on $B$. We write $\varPhi^*$ for
the Fourier Mukai transfrom with respect to $\Pc^*$.

Further let us write $\tau\colon X\to X$ for the elliptic involution on
$X\to B$ so that we write $\tau^*F$ instead of $F^\tau$ for $F$ in the
derived category. We then have three involutions on $X\times_B X$:
\begin{align*}
\tau(x,y)&=(\tau(x),y)=(-x,y)\\
\hat\tau(x,y)&=(x,\tau(y))=(x,-y) \\
\bar\tau&=\hat\tau\circ\tau=\tau\circ\hat\tau.
\end{align*}
One easily sees that $\hat\tau^*\Pc=\Pc^*$ and one has
\begin{align*}
\varPhi(\tau^* F)&=\pi_{2,*}(\pi_{1}^*(\tau^* F\otimes
\Pc))=
\pi_{2,*}(\tau^*(\pi_{1}^* F\otimes \tau^*\Pc))
\\
&=\pi_{2,*}(\hat\tau^*\bar\tau^*(\pi_{1}^* F\otimes \tau^*\Pc))
=\hat\tau^*\pi_{2,*}(\bar\tau^*(\pi_{1}^* F\otimes \tau^*\Pc))
\\
&= \hat\tau^*\pi_{2,*}(\pi_{1}^* (\tau^* F)\otimes \hat\tau^*\Pc))
= \hat\tau^*\pi_{2,*}(\pi_{1}^* \tau^* F\otimes \Pc^*)
=  \hat\tau^* \varPhi^*(\tau^*F)\,.
\end{align*}
Now if $\tau^*F=F$, one has
\begin{equation}
\varPhi(F)=\hat\tau^* \varPhi^*(F)\,.
\label{fminvsec}
\end{equation}
If we assume that $F$ reduces to a single stable, irreducible,
holomorphic $SU(n)$ vector bundle over elliptic Calabi-Yau specified by
a pair $(C,L)$ via the inverse Fourier Mukai transform $\hat \varPhi$. Then
$\varPhi(F)$ reduce to the sheaf $\varPhi^1(F)=i_*(L)$, where $i\colon C\hookrightarrow X$ is
the immersion and \ref{fminvsec} means that $\varPhi^*(F)$ reduces to a single
sheaf $\varPhi^{*1}(F)$ as well and that $\varPhi^{*1}(F)=\hat\tau^* (i_*(L))$. If
$C^\tau=\tau(C)$, $L^\tau=\tau^*L$ and $j\colon C^\tau\to X$ is the
immersion, it follows that
$$
 \varPhi^{*1}(F)=j_*(L^\tau)\,.
 $$
On can say that if $\tau^* F=F$, then $F$ can be specified by spectral
data in two different ways, either by $(C,L)$ via the inverse Fourier
Mukai transform $\hat \varPhi$ or by $(C^\tau,L^\tau\otimes (q^*K_B)_{|C})$
via the standard FM (with respect to $\Pc$).

\subsubsection*{Moduli of $\pi^*M$}\

\smallskip

The number of moduli of $\pi^*M$ are given by $h^1(X, \pi^*\End(M))$. If
one applies the Leray spectral sequence to the elliptic fibration of
$X$ and
assumes that the moduli space of $M$ over $B$ is smooth then one can
show \cite{BJPS} that
all moduli of $\pi^*M$ come from moduli of $M$ on the base $B$. The
moduli can be then
evaluated using the Riemann-Roch index theorem (assuming $M$ being a
$SU(m)$
bundle and $B$ a rational surface)
$$
h^1(B, \End(M))=2mk-(m^2-1)\,.
$$

\subsubsection*{What is Known About the Remaining
Moduli?}\ 

\smallskip

First we note that
\begin{align*}
H^i(\Hom(V,\pi^*M))&=\Ext^i(V,\pi^*M) \\
H^i(\Hom(\pi^*M,V))&=\Ext^i(\pi^*M,V)
\end{align*}
since $V$ and $\pi^*M$ are locally free sheaves on $X$. In particular,
elements of the vector spaces $H^1(\Hom(V,\pi^*M))$ and
$H^1(\Hom(\pi^*M,V))$ give non-trivial extensions $0\rightarrow
\pi^*M\rightarrow E_{\mu}\rightarrow V\rightarrow0$ respectively
$0\rightarrow V \rightarrow E^{\nu}\rightarrow \pi^*M\rightarrow 0$.
More information can be obtained by computing the
index
\begin{equation}
I_X=\sum_{i=0}^{3}(-1)^i \dim H^i(\Hom(V,\pi^*M))\,.
\label{fiid}
\end{equation}
We will later show that $\dim H^i(\Hom(V,\pi^*M))=0$ for $i=0,3$.
Further we note that $H^i(\Hom(V,\pi^*M))=H^i(V^*\otimes \pi^*M)$ and
$H^i(\Hom(\pi^*M,V))=H^i((\pi^*M)^*\otimes V)$
and applying Serre duality we get $H^2(V^*\otimes
\pi^*M)=H^{1}(V\otimes (\pi^*M)^*)$ using the fact
that the canonical bundle $K_X$ of $X$ is trivial. Thus we get
\begin{equation}
\begin{aligned}
I_X&=\dim H^1(\Hom(\pi^*M,V))-\dim H^1(\Hom(V,\pi^*M)) \\
& = \dim \Ext^1(\pi^*M,V))-\dim \Ext^1(V,\pi^*M)
\end{aligned}
\label{inda}
\end{equation}
The left hand side of \ref{fiid} can be evaluated using the Riemann-Roch
theorem
\begin{equation}
I_X=\int_X ch(V^*)ch(\pi^*M)\op{Td}(X)=- \frac 12 mc_3(V)
\label{lhs}
\end{equation}
and is related to chiral matter for non-zero $c_3(V)$ as observed in
\cite{BJPS}. The computation of $c_3(V)$ in the spectral cover and
the parabolic bundle construction has been performed in \cite{C}
respectively \cite{And}.

\subsection{Evaluation of the Remaining Moduli}\ 

\medskip
We will now proceed as follows. We first rewrite the index in terms of
the spectral data using the so-called Parseval theorem and then
restrict to
$S$ to further evaluate the index.

The Parseval theorem for the relative Fourier-Mukai transform has been
proved by Mukai in his original Fourier-Mukai transform for abelian
varieties \cite{Muk}, but can be easily extended to any situation in which a
Fourier-Mukai transform is an equivalence of categories.

\bigskip\noindent{\it Parseval Theorem}\smallskip\nobreak

Assume that we have sheaves ${\F}$, $\bar {\F}$ that are
respectively WIT$_h$ and WIT$_j$ for certain $h$, $j$; this means  that
they only have one non-vanishing Fourier-Mukai transform, the $h$-th
one  ${\varPhi}  ^h({\F})$ in the case of ${\F}$ and the
$j$-th one ${\varPhi}  ^j(\bar {\F})$ in
the case of $\bar {\F}$. Parseval theorem says that one has
\begin{equation}
\Ext_X^i({\F},\bar {\F})=\Ext_X^{h-j+i}({\varPhi}
^h({\F}),{\varPhi}  ^j(\bar {\F}))\,,
\label{preparc}
\end{equation}
thus giving then a correspondence between the extensions of ${\F}$,
$\bar{\F}$ and the extensions of their Fourier-Mukai transforms.
The proof is very simple, and relays on two facts. The first one is
that for arbitrary coherent sheaves $E$, $G$ the ext-groups can be
computed in terms of the derived category, namely
\begin{equation}
\Ext^i(E,G)=\Hom_{D(X)}(E,G[i])
\label{extdef}
\end{equation}
The second one is  that  the Fourier-Mukai transforms of
${\F}$, $\bar {\F}$ in the derived category $D(X)$ are
${\varPhi}  ({\F})={\varPhi}  ^h({\F})[-h]$,
${\varPhi}  (\bar {\F})={\varPhi}  ^j(\bar {\F})[-j]$.
Now, since the Fourier-Mukai transform is an equivalence of categories, one has
\begin{align*}
\Hom_{D(X)}({\F},\bar {\F}[i]) & =Hom_{D(X)}({\varPhi}  ({\F}),{\varPhi}  (\bar
{\F}[i])) \\
&=\Hom_{D(X)}({\varPhi}  ^h({\F})[-h],{\varPhi}  ^j(\bar {\F})[-j+i]) \\
&=Hom_{D(X)}({\varPhi}  ^h({\F}),{\varPhi}  ^j(\bar {\F})[h-j+i])
\end{align*}
so that \ref{extdef} gives the Parseval theorem \ref{preparc}.

In particular, if both ${\F}$ and $\bar {\F}$ are WIT$_j$ for
the same $j$,
we obtain
$$
\Ext_X^i({\F},\bar {\F})=\Ext_X^i({\varPhi}
^j({\F}),{\varPhi}  ^j(\bar {\F}))
$$
for every $i\ge 0$.

We can apply Parseval theorem to our situation, because $V$ and $\pi^*M$
are WIT$_1$. Since their Fourier-Mukai transforms are respectively
$i_*L$ where
$i\colon C\hookrightarrow X$ is the immersion, and $\sigma_*(G)$ where
$\sigma\colon B\to X$ is the
section and $G=M(K_B)$ \ref{gsheaf}, we have
\begin{equation}
\Ext_X^i(V,\pi^*M)=\Ext_X^i(i_*L,\sigma_*(G))\,.\quad i\geq 0
\label{parc}
\end{equation}
Now we note that, due to the fact that $V$ and $\pi^*M$ are vector
bundles, we have $H^1(\Hom(V,\pi^*M))=\Ext^1_X(V,\pi^*M)$ and
$H^1(\Hom(\pi^*M, V))^*=\Ext^2_X(V,\pi^*M)$; as we are computing
dimensions
we have $\dim H^1(\Hom(\pi^*M, V))^*=\dim H^1(\Hom(\pi^*M, V))$ so that
we can
actually rewrite the index $I_X$ in terms of the spectral bundles
$$
I_X=\sum_{i=0}^3 (-1)^i \dim\Ext^i_X(i_*L,\sigma_*(G))\,.
$$

\subsubsection*{Restriction to $S$}\ 

\smallskip

We now proceed as in Section 6 of \cite{OUR} and use the sequence of low
terms of the spectral sequence associated to Grothendieck duality for
the immersion $i$.

In its simpler form, Grothendieck duality for a smooth morphism is a
sort of relative Serre duality, a Serre duality for flat families of
smooth varieties. Grothendieck extended this notion to very general
morphisms of algebraic varieties; his formulation requires derived
categories and a notion of ``dualizing complex''; this is an object of
the derived category that plays (for a general morphism) the same role
as
the sheaf of $r$-forms on a smooth $r$-dimensional variety.

We now apply  Grothendieck duality  for the closed
immersion $i\colon C\hookrightarrow X$ and denote by
$\tilde\sigma$ the
restriction of  $\sigma$ to $S$.  Now Grothendieck duality
for the closed immersion $i\colon C\hookrightarrow
X$ says that there is an isomorphism in the derived
category
$$
R\, \Hom_X(i_*L,\sigma_*G)= R\, \Hom_C(L,i^!(\sigma_*G))
$$
where $i^!(\sigma_*G)$ is the ``dualizing complex'' for the immersion,
and $i^!(\sigma_*G)$ is determined by the equation
$i_*(i^!(\sigma_*G))=R\,\Hc om_{{\Oc}_X}(i_*{\Oc}_C,\sigma_*G)$ where $\Hc om$ stands for the Hom-sheaf (see \cite{RD}
Section \S 6). Now let us consider the exact sequence
\begin{equation}
0\to {\Oc}_X(-C)\to{\Oc}_X\to i_*{\Oc}_C\to 0
\label{esaw}
\end{equation}
where ${\Oc}_X$ and ${\Oc}_C$ are the trivial bundles (structure
sheaves) on $X$ and $C$;
${\Oc}_X(-C)$ is the inverse of the tautologically defined line
bundle ${\Oc}_X(C)$
on $X$ that admits a holomorphic section $s$ that vanishes precisely on
$C$; also note
the first map in \ref{esaw} is multiplication by $s$ and the second is
restriction to $C$. We need $i_*$ to understand ${\Oc}_C$ as a sheaf
on $X$, the sheaf that coincides with ${\Oc}_C$ on $C$ and it is zero
on $X-C$.

 From \ref{esaw} we read that $R \Hc om_{{\Oc}_X}(i_*{\Oc}_C,\sigma_*G)$ is represented by the complex
$$
\sigma_*G\xrightarrow{d=0}  \Hc om_{{\Oc}_X}({\Oc}_X(-C),\sigma_*G)
$$
that is,
$$
i^!(\sigma_*G) = \{\, \tilde\sigma_*(G_{\vert S})\xrightarrow{d=0}
 \tilde\sigma_*(G_{\vert S}\otimes (N_{X/C})_{\vert S})\,\} 
$$
in the derived category, where $N_{X/C}$ is the normal sheaf to $C$ in
$X$. Then, since $L$ is a line bundle, we have
\begin{align*}
R\, \Hom_X(i_*L,\! \sigma_*G)&=R\, \Hom_C(L, \{ \,\tilde\sigma_*(G_{\vert S})\xrightarrow{d=0} \tilde\sigma_*(G_{\vert S}\otimes
(N_{X/C})_{\vert S})\,\} ) \\
&=R\,\Gamma(C, \{\,  L^{-1}\otimes
\tilde\sigma_*(G_{\vert S}) \xrightarrow{d=0}
L^{-1}\otimes\tilde\sigma_*(G_{\vert S}\otimes (N_{X/C})_{\vert
S})\,\})\,.
\end{align*}
The  above equality in the derived category means that we can  approach
the
cohomology groups
$\Ext_X^i(i_*L,\sigma_*G)$ on the left hand side from a double complex of
global sections of an acyclic resolution of $ \{\,  L^{-1}\otimes
\tilde\sigma_*(G_{\vert S}) \xrightarrow{d=0}
L^{-1}\otimes\tilde\sigma_*(G_{\vert S}\otimes (N_{X/C})_{\vert
S})\,\}$. We have
\begin{align*}
E_2^{p,0} & = \Ext^p_C(L,\tilde\sigma_*(G_{\vert S})) \\
E_2^{p,1} & = \Ext^p_C(L, \tilde\sigma_*(G_{\vert S}\otimes
(N_{X/C})_{\vert S}))  \\
E_2^{p,q} & = 0\,,\qquad \text{for  $q>1$}\,.
\end{align*}
We have the exact sequence of the low terms
$$
0\to E_2^{1,0}\to
H^1(M)\to E_2^{0,1} \xrightarrow{d_2} E_2^{2,0}\to
H^2(M)\,.
$$
Moreover, the spectral sequence is of ``spherical fiber'' type, that is,
one has $E_2^{p,q}=0$ for every $p$ and $q\neq 0,1$\footnote{Spectral
sequences with $E_2^{p,q}=0$ for every $p$ and $q\neq 0,m$ are called
of ``spherical fiber'' type, because when one has a bundle $Z\to Y$ on
$m$-spheres, the Leray spectral sequence approaching the cohomology of
$Z$ in terms of the cohomology of $X$ is of that kind.}.
It is a standard fact  (see for
instance \cite{Go}  paragraph 4.6), that we can complete the above sequence
to get
\begin{multline*}
0\to E_2^{1,0}\to H^1(M)\to E_2^{0,1}\xrightarrow{d_2}
E_2^{2,0}\to H^2(M) \to  \\
\to E_2^{1,1} \xrightarrow{d_2}
E_2^{3,0}\to H^3(M) \to E_2^{2,1} \xrightarrow{d_2}
E_2^{4,0}.
\end{multline*}
Moreover,  since $d=0$ the first differential of this double complex is
zero and then $d_2$ is zero as well.  This implies that the above exact
sequence breaks into short exact sequences. We then have
$$
0\to \Ext^1_C(L,\tilde\sigma_*(G_{\vert S})) \to
\Ext_X^1(i_*L,\sigma_*(G))
\to \Hom_C(L, \tilde\sigma_*(G_{\vert S}\otimes (N_{X/C})_{\vert S}))
\to 0
$$
and  isomorphisms
\begin{align*}
\Ext_X^2(i_*L,\sigma_*(G))
&\simeq \Ext_C^1(L, \tilde\sigma_*(G_{\vert S}\otimes (N_{X/C})_{\vert
S})),
\\
\Ext_X^3(i_*L,\sigma_*(G))
&\simeq \Ext_C^2(L, \tilde\sigma_*(G_{\vert S}\otimes (N_{X/C})_{\vert
S}))
\end{align*}
due to the fact that  $\Ext^i_C(L,\tilde\sigma_*(G_{\vert S}))=H^i(S,
L^{-1}_{\vert S}\otimes G_{\vert S})=0$ for $i\ge 2$.  But we have
$\Ext_C^2(L, \tilde\sigma_*(G_{\vert S}\otimes (N_{X/C})_{\vert
S}))=H^2(S, L^{-1}_{\vert S}\otimes G_{\vert S}\otimes (N_{X/C})_{\vert
S})=0$ as well, so that
\begin{equation}
\Ext_X^3(i_*L,\sigma_*(G))=0.
\label{hthree}
\end{equation}
On the other hand
\begin{equation}
\Hom_X(i_*L,\sigma_*(G))=\Hom_B(\sigma^*(i_*L), G) =0
\label{hnot}
\end{equation}
because $\sigma^*(i_*L)$ is concentrated on $S$ and $G$ is a vector
bundle. In the following we set
$$
{\F}=L^{-1}_{\vert S}\otimes G_{\vert S}\otimes
(N_{X/C})_{\vert S}\,.
$$

We have
\begin{align*}
\Hom_C(L, \tilde\sigma_*(G_{\vert S}\otimes (N_{X/C})_{\vert S})) &=
H^0(S,{\F})\\
\Ext_C^1(L, \tilde\sigma_*(G_{\vert S}\otimes (N_{X/C})_{\vert S})) &=
H^1(S,{\F})
\end{align*}
where $N_{C/S}$ is the normal bundle to $S$ in $C$. Further we set
$\dim H^i(S,{\F})=h^i(S,{\F})$ as usual. Now from \ref{hthree} and
\ref{hnot} we find that the index $I_X$ simplifies to
$$
I_X=\sum_{i=1}^2 (-1)^i\dim \Ext^i_X(i_*L,\sigma_*(G))
$$
in agreement with \ref{inda}. Thus we get for the dimensions we were looking
for
\begin{equation}
\begin{aligned}
\dim \Ext^1_X(i_*L,\sigma_*(G))&=-I_X+h^1(S,{\F})\,,\\
\dim \Ext^2_X(i_*L,\sigma_*(G))&=h^1(S,{\F})\,.
\end{aligned}
\label{dims}
\end{equation}
Since we know that the value of the index $I_X=-\frac12 mc_3(V)$ by
\ref{lhs},
we have reduced the computation of the dimensions of
$H^1(\Hom(V,\pi^*M))$ and $H^1(\Hom(
\pi^*M, V))$ to the computation of the dimension of
the first cohomology group of the vector bundle ${\F}$ on the
intersection
curve $S$ of $C$ and $\sigma$. Thus the question remains: can we
actually compute $h^1(S,{\F})$?
Let us assume that $S$ is irreducible.
Something we can do is to compute another index, namely, we can use the
Riemann-Roch theorem on $S$ to compute
$$
\chi(S,{\F})=h^0(S,{\F})-h^1(S,{\F})\,.
$$
When $S$ is smooth this index can be computed as $\int_S ch({\F})\op{Td}(S)$.
As we are working on $S$, the computation is reduced to the evaluation
of the first Chern-class of ${\F}$. The details of this computation
are given in appendix C, the result is
$$
c_1({\F})= \frac12 m(3C\sigma^2+C^2\sigma) +  \frac12 m c_3(V)\,.
$$
Since $S=C\cdot \sigma$ and we are assuming that $C$ is a Cartier
divisor in $X$, it follows that $S$ is a Cartier divisor in $B$ so that it is a
Gorenstein curve. This means that it has a canonical divisor $K_S$
with all properties that the usual canonical divisor for a
smooth curve has. In particular we have both the Riemann-Roch theorem and Serre
duality for $S$. Now the Riemann-Roch theorem for the curve $S$ gives
$$
h^1(S,{\F})= h^0(S,{\F})-mC\sigma^2-\frac12 m c_3(V)
$$
which reduces the problem either to compute the number of sections of
${\F}$ or, by Serre duality, to compute the number of sections of
$$
{\F}^{\vee}={\F}^*\otimes K_S=L_{\vert S}\otimes
M^{-1}_{\vert S}\,.
$$
In the next section we will give some vanishing arguments
such that ${\F}^{\vee}$ has no sections.

\subsection{Applying a Vanishing Theorem}\ 

\medskip
We are always assuming in this subsection that our curve $S$ has
arithmetic genus $g(S)=h^1(S,{\Oc}_S)>1$. When $S$ is smooth, the
genus of $S$ is determined from the known formula
$e(S)=2-2g(S)$ where $e(S)$ denotes the topological Euler
characteristic of $S$.
To compute $g(S)$ one considers first the canonical bundles of $C$ and
$B$
in $X$
\begin{align*}
K_C&={K_X}_{\vert C}+N_{X/C}=N_{X/C}\\
K_B&={K_X}_{\vert B}+N_{X/B}=N_{X/B}
\end{align*}
where $N_{X/B}$ respectively $N_{X/C}$ denotes the normal bundles
of $C$ and $B$ in $X$. One then considers the canonical divisor
$K_S={K_C}_{\vert S}+N_{C/S}$
or equivalently $K_S={K_B}_{\vert S}+N_{B/S}$ with $N_{B/S}=C^2\sigma$
and
$N_{C/S}=C\sigma^2$ and finds
\begin{equation}
2g(S)-2=C^2\sigma + C\sigma^2
\label{gS}
\end{equation}
Now in general, a vector bundle on a projective variety has no sections
if it is stable of negative degree.  As we are working  on a curve $S$,
to
prove that  a stable vector bundle $\bar{\F}$ of rank greater than
1 on $S$ has no
sections, we
need $c_1(\bar{\F})\le 0$ because a section gives a trivial
subbundle ${\Oc}\subset \bar{\F}$.\footnote{For a line bundle $L$,
conditions $c_1(L)\le 0$ implies that $L$ has no sections except if $L$
is trivial}

In some cases we can have generically the vanishing of the sections
without having negative first Chern class. This happens, always for
stable
$\bar{\F}$, when one has $\chi(S,\bar{\F})\leq 0$, because those
sheaves which have sections define the so called $\Theta$-divisor in
the moduli
space of such sheaves; this means that we can deform
$\bar{\F}$ to a stable sheaf with no sections. But in our
situation, we will actually find that $\bar{\F}$ is stable and
$\chi(S,\bar{\F})> 0$, and then the above genericity argument does not apply; we need
$c_1(\bar{\F})\le 0$ to ensure that  $\bar{\F}$ has no sections.

We want to prove that ${\F}^{\vee}$ has no sections. Before
considering the stability question, we just make sure that $c_1({\F}^{\vee})\le 0$,
which together with stability gives the desired vanishing theorem.

We first recall that due to \ref{gS},  the
condition $g(S)>1$ is equivalent to
$$
C^2\sigma+C\sigma^2 >0\,.
$$
Further we have
\begin{equation} mC^2\sigma \le \frac12 m c_3(V)
\label{cthree}
\end{equation}
since $\frac12 m c_3(V)
-mC^2\sigma=h^0(S,L^{-1}_{\vert S}\otimes G_{\vert S})\ge 0$ (see
appendix D). Then
\begin{equation}
c_1({\F}^{\vee})=mc_1(L_{\vert S})=\frac12 m
(C^2\sigma-C\sigma^2)-
\frac12 mc_3(V)< m C^2\sigma -\frac12 m c_3(V)\le 0
\label{neg}
\end{equation}
as claimed (the expression for $c_1(L_{\vert S})$ is given in the
appendix C).

We are now  going to see that we can select our bundle $M$ on $B$ in
such a way that the restriction of $M$ to $S$ is stable\footnote{We would
like to thank
G. Hein for helpful discussions!}. Since the dual of a
stable vector bundle is stable and twisting a vector bundle by a line
bundle does not
affect the stability. The sheaf ${\F}^{\vee}$ will be stable as
well.
For this we recall that on a curve of $g>1$ the
general deformation of a vector bundle is stable \cite{RF}. Thus we need
to make sure that $M_{\vert S}$, as object in the local moduli space
$\operatorname{Def}(M_{\vert S})$ of bundles on $S$, can be deformed in arbitrary
directions in its moduli space if we deform $M$ in its local moduli
space $\operatorname{Def}(M)$. Then we want that the restriction map $\tau\colon
\operatorname{Def}(M_{\vert
S})\to \operatorname{Def}(M)$ be surjective (or more technically, that the map
defined between the local deformation functors be surjective). Let us
consider
the exact sequence:
$$
0\rightarrow \op{ad} M\otimes {\Oc}(-S)\rightarrow
\op{ad} M\rightarrow
\op{ad} M_{\vert S}\rightarrow 0
$$
where $\op{ad}M$ are the traceless endomorphisms of $M$. This gives
rise to
a long exact sequence
\begin{align*}
\rightarrow H^1(\op{ad} M) \xrightarrow{d\tau} H^1(\op{ad} M_{\vert S})
\rightarrow & H^2(\op{ad} M\otimes {\Oc}_B(-S))\rightarrow \\
& H^2(\op{ad} M\otimes {\Oc}_B(-S)) \to H^2(\op{ad}
M)\to 0\,.
\end{align*}
Thus if $H^2(\op{ad} M\otimes {\Oc}_B(-S))=0$ then $H^2(\op{ad}
M)=0$ so that $\tau$ is surjective and deformations
of $M$ give a general deformation of $M_{\vert S}$  \cite{RF}. Serre duality
on
$S$ gives $H^2(\op{ad} M\otimes {\Oc}(-S))=H^0(K_B\otimes \op{ad}
M\otimes {\Oc}(S))$, and then the above condition transforms to
\begin{equation}
H^0(K_B\otimes \op{ad} M\otimes {\Oc}(S))=0\,.
\label{see}
\end{equation}

Now, whenever \ref{see} is satisfied, we can deform $M$ so that $M_{\vert
S}$
is stable and thus ${\F}^{\vee}$ is stable as well. Since we already
know by \ref{neg} that $c_1({\F}^{\vee})<0$, we get $h^0(S,{\F}^{\vee})=0$. By Serre duality, 
\begin{align*}
h^1(S,{\F})&=h^0(S,{\F}^{\vee})=0 \\
h^0(S,{\F})&=\chi(S,{\F})=m C\sigma^2 +\frac12 mc_3(V)
\end{align*}
and then  we can compute directly, via the Parseval equality \ref{parc},
equation
\ref{dims} and the computation \ref{parc} of the index $I_X$, the number of
moduli we were looking for:
\begin{equation}\begin{aligned}
h^1(\Hom(V,\pi^*M))& =\dim \Ext^1_X(V,\pi^*M) = -I_X+ h^1(S,{\F})=\frac12 mc_3(V) \\
h^1(\Hom(\pi^*M,V))& =\dim \Ext^2_X(V,\pi^*M) =h^1(S,{\F})= 0\,.
\end{aligned}
\label{lejttwo}
\end{equation}

\subsubsection*{Conditions for the Vanishing of $H^0(K_B\otimes
\op{ad}M\otimes {\Oc}(S))$:}\ 

\smallskip

Now we consider two situations where \ref{see} is true and we can
apply the vanishing theorem to get the number of moduli \ref{lejttwo}.

\subsubsection*{Case 1: Conditions on the Curve}\ 

\smallskip

Since $\op{ad} M$ is stable with $c_1(\op{ad} M)=0$, we get
$H^0(K_B\otimes \op{ad} M\otimes {\Oc}(S))=0$ if
$$
\op{deg}(K_B\otimes{\Oc}(S))\leq 0 
$$
is satisfied for an arbitrary ample $H$ in $H^2(B)$. 

\subsubsection*{Case 2: Conditions on the Bundle
$V$}\ 

\smallskip

By Theorem 40 of \cite{RF}, there is a constant $k_0$ (depending on $B$, the
polarization considered in $B$ and the curve $S$), such that for
$c_2(M)=k\ge k_0$ the vanishing equation \ref{see} is true.\footnote{Theorem
41 of \cite{RF} can be applied as well to see directly that $M$ can be deformed to
have $M_{\vert S}$ stable on $S$} We are not completely free to choose
$k$,
because $k$ is related to $V$ since is the number of vertical branes we
wanted to remove. That is, we are constrained to have
$$
k\le a_F
$$
where $a_F$ is the number of fibers contained in the class $[W]=c_2(TX)-
c_2(V)$ (see also \cite{OPP}). If we write $c_2(V)=\sigma\pi^*(\eta)
+\pi^*(\omega)$, we have
$$
a_F= \int c_2(B)-c_1(B)^2-\omega
$$
Since the base surface $B$ is fixed, the Chern classes $c_i(B)$ are
fixed as well, and we see that $a_F$ just depends on the choice of
$\omega$.

Now if we write $a=-\int_B\omega$, using (2.32) of \cite{OUR}, we have
$$
a=- \frac16 n c_1(B)^2+ch_3(i_*L)
$$
(because ${\varPhi}^1(V)=i_*L$), and Grothendieck Riemann-Roch gives
$$
a= - \frac16 n c_1(B)^2+\frac12 c_1(L)(c_1(L)-i^*C)+\frac1{12}
(i^*C)^2
$$
where the intersections are made inside $C$.

We then proceed in this way: we fix the spectral cover $C$, so that
$S=C\sigma$
is  an irreducible curve of arithmetic genus $g(S)>1$. Then we take the 
corresponding aforementioned
constant $k_0$, and an arbitrary $k\ge k_0$. We can then take a line
bundle $L$
on $C$ so that $c_1(L)(c_1(L)-i^*C)$ is big enough to have $k\le a_F$.
This can be done as follows:  take $L'$ very ample and $q\gg 0$ in such
a
way that $qc_1(L')-i^*C$ is a  very ample divisor; if $L=(L')^{\otimes
q}$, then
$c_1(L)(c_1(L)-i^*C)$ grows as $q^2$ so that for $q\gg0$ $L$ fulfills
our requirements.

As $V=\hat{\varPhi}^0(i_*L)$ is a vector bundle on $X$ with
spectral cover $C$ and in this situation we can take a stable bundle
$M$ on $B$ with $rk(M)=m$, $c_1(M)=0$ and $c_2(M)$ so that $M_{\vert
S}$ is stable and we have the vanishing theorem and the formulas
\ref{lejttwo}
for the number of moduli.

\section{Comments on Localized Chiral Matter}

In this section we will analyze the question: can we determine not only
the net amount of chiral matter but also the matter multiplets
individually? Let us first motivate this question from various
perspectives.

As it is well known, the net amount of chiral matter is determined in
heterotic string compactification on Calabi-Yau threefolds by $\frac12 c_3(V)$.
This follows from the fact that chiral fermions in four dimensions are
related to
zero modes of the Dirac operator on $X$. The index can be written as
(denoting as usual
$h^i(X,V)=\dim H^i(X,V)$)
$$
\op{index}(D_V)=\sum_{i=0}^3(-1)^ih^i(X,V)=\int_X ch(V)\op{Td}(X)
$$
and for a stable bundle $V$ with $c_1(V)=0$ we have
$h^0(X,V)=h^3(X,V)=0$, then
$$
-(h^1(X,V)-h^2(X,V))=\frac12 c_3(V)\,.
$$
Now for $V$ an $SU(n)$ vector bundle the corresponding unbroken
space-time gauge group is the maximal subgroup of $E_8$ which commutes
with $SU(n)$. For instance taking $V$ a general $SU(3)$ bundle the unbroken
observable
gauge group is $E_6$. The only charged ten-dimensional fermions are in
the adjoint
representation of $E_8$ thus we get four-dimensional fermions only from
the reduction
of the adjoint representation. In particular one has under $E_6\times
SU(3)$
$$
{\bf 248}=({\bf 78},{\bf 1})+({\bf 27},{\bf
3})+({\bf\overline{27}},{\bf\bar 3})+({\bf 1},{\bf 8})\,.
$$
Therefore fermions that are in the ${\bf 27}$ of $E_6$ are in the ${\bf
3}$ of $SU(3)$ thus
in the index the left handed ${\bf 27}$'s can be assigned\footnote{Note the assignment is a matter of convention. In case of $V=TX$ one typically assigns the ${\bf 27}$'s to elements of $H^1(X,V)$
as the Euler characteristic in typical examples turns out to be negative.} to elements of
$H^2(X,V)=H^1(X,V^*)$ and
the left handed ${\bf\overline{27}}$'s would be assigned to elements of
$H^1(X,V)$.
In case of $V$ being the tangent  bundle $TX$ one simply has
$h^1(X,TX)=h^{1,2}(X)$ and  $h^2(X,TX)=h^{1,1}(X)$.
 In addition one can analyze
\cite{Witnew} the corresponding Yukawa couplings taking into account the
number of tangent bundle moduli $h^1(X, \End(TX))$ associated to
$E_6$ singlets. The resulting Yukawa couplings are: ${\bf 27}^3$,
${\bf\overline{27}}^3$, ${\bf 27}\cdot{\bf\overline{27}}\cdot{\bf 1}$,
${\bf 1}^3$.  Now assuming that $h^1(X,\End(V))=0$
one would expect the ${\bf 27}^3$, ${\bf\overline{27}}^3$ terms only
\cite{Witnew}. Similarly if $H^1(X,V)$ or $H^1(X,V^*)$ would vanish one would
expect the vanishing of the corresponding couplings.

We will now proceed as in section 4 to compute each of the
groups\footnote{The localization
of $H^i(X,V)$ was originally suggested in \cite{FMW} and worked out
precisely in \cite{DI} using the
Leray spectral sequence. We give here an alternative approach and
extend the discussion
by giving a vanishing argument similar to the one discussed in the
previous section.}
$H^i(X,V)$ and their dimensions, just by taking the bundle $M$ as the
trivial line
bundle ${\Oc}_B$. We have
\begin{equation}
H^i(X,V)=\Ext^i_X({\Oc}_X,V)=\Ext^{3-i}_X(V,{\Oc}_X)
\label{duality}
\end{equation}
so that the index $\Ic=\sum_{i=0}^3(-1)^i\dim \Ext^i(V,{\Oc}_X)$
fulfills
$$
\Ic=-\op{index}(D_V)=- \frac12 c_3(V)
$$
We also have a Parseval equation like \ref{parc}
\begin{equation}
\Ext_X^i(V,{\Oc}_X)=\Ext_X^i(i_*L,\sigma_*(K_B))
\end{equation}
and the new equation that corresponds to \ref{dims} is
\begin{equation}
\begin{aligned}
\dim \Ext^1_X(i_*L,\sigma_*(K_B))&=- \Ic+h^1(S,{\F}') \\
\dim \Ext^2_X(i_*L,\sigma_*(K_B))&=h^1(S,{\F}')
\end{aligned}
\label{newdims}
\end{equation}
with ${\F}'=L^{-1}_{\vert S}\otimes K_S$.
By Serre duality on $S$, $h^1(S,{\F}')=h^0(S,L_{\vert S})$; then,
\ref{duality} and \ref{newdims} lead to
\begin{align*}
h^2(X,V)=\dim \Ext_X^1(V,{\Oc}_X)= \dim
\Ext^1_X(i_*L,\sigma_*(K_B))&=-
\Ic +h^0(S,L_{\vert S}) \\
h^1(X,V)= \dim \Ext_X^2(V,{\Oc}_X)=\dim
\Ext^2_X(i_*L,\sigma_*(K_B))&=h^0(S,L_{\vert S})\,.
\end{align*}
Now we have, by Appendix C,
$$
c_1(L_{\vert S})= \frac12(- \sigma^2 C+\sigma
C^2)-\frac12  c_3(V)=
1-g(S)+\sigma C^2- \frac12 c_3(V)\,.
$$
If we assume that $g(S)> 1$,  and since $C^2\sigma- \frac12 c_3(V)\le
0$ by
\ref{cthree}, we have
$$
c_1(L_{\vert S})<0\,.
$$
It follows that $H^0(S,L_{\vert S})=0$. Thus we see that
\begin{align*}
h^2(X,V)&=- \Ic = \op{index}(D_V)=\frac12 c_3(V)\\
h^1(X,V)&= 0\,.
\end{align*}
We conclude that chiral matter associated to
$H^1(X,V)$ vanishes in heterotic string compactifications on
elliptically fibered $X$
with vector bundles constructed in the spectral cover approach and
matter
localized along the curve $S$ of genus $g(S)>1$. This implies the
vanishing of
the corresponding Yukawa couplings. For example, we would expect for
$V=SU(3)$
the Yukawa couplings: ${\bf{27}}^3$ and ${\bf 1}^3$.

\subsection*{Acknowledgments} We would like to thank G. Hein, H. Kurke and E. Witten for helpful
comments.

\appendix

\section{Hilbert Polynomial Coefficients}

In this appendix we provide the coefficients required for section $2.3$.
The coefficients of  $P({\F},m)$ are given by
\begin{align*}
\alpha_0({\F})&=\ch_3({\F})+\ch_1({\F}) \frac{c_2(TX)}{12}\\
\alpha_1({\F})&=\ch_2({\F})\tilde{H}+\ch_0({\F})\frac{c_2(TX)}{12}\tilde{H}\\
\alpha_2({\F})&=\ch_1({\F})\tilde{H}^2\\
\alpha_3({\F})&=\ch_0({\F})\tilde{H}^3\,.
\end{align*}

The coefficients of $P(\hat{{\varPhi}  }(V),m)$ are given by
\begin{align*}
\alpha_1(\hat{{\varPhi}  }(V))&=(( \frac12 nc_1+\tilde
S)\sigma-(s+ \frac12 \eta c_1\sigma)F) \tilde{H}\\
\alpha_2(\hat{{\varPhi}  }(V))&=(n\sigma-\eta)\tilde{H}^2\,.
\end{align*}

The coefficients of $P(\hat{{\varPhi}  }(V'),m)$ are given by
\begin{align*}
\alpha_1(\hat{{\varPhi}  }(V'))&=(( \frac12 n'c_1+\tilde
S')\sigma-(s'+ \frac12 \eta' c_1\sigma)F) \tilde{H}\\
\alpha_2(\hat{{\varPhi}  }(V'))&=(n'\sigma-\eta')\tilde{H}^2\,.
\end{align*}

\section{A Simple Test}

We want to compute the index $I_X=\sum_{i=0}^3(-1)^i\dim
\Ext^i_X(i_*L,\sigma_*(G))$.
The Riemann-Roch theorem gives
\begin{equation}
\begin{aligned} 
I_X&=\int_X \ch(i_*L^{-1})\ch(\sigma_*(G))\op{Td} (X)\\
&=\int_X
\ch_1(i_*L^{-1})\ch_2(\sigma_*(G))+\ch_1(\sigma_*(G))\ch_2(i_*L^{-1})
\end{aligned}
\label{ind}
\end{equation}
The relevant Chern characters of $i_*L$ are given by \cite{OUR}
\begin{align*}
\ch_1(i_*L^{-1})&=n\sigma+\eta\\
\ch_2(i_*L^{-1})&= \frac12 nc_1(B)\sigma-(\ch_3(V)- \frac 12 \eta
c_1(B)\sigma)\cdot F
\end{align*}
where $F$ denotes the class of the fiber of $\pi\colon X\rightarrow B$.
Further we have to
obtain the relevant Chern characters of $\sigma_*(G)$. We can compute
these using
the Grothendieck-Riemann-Roch theorem for $\sigma\colon B\rightarrow X$,
which is $\ch(\sigma_*G)\op{Td}(X)=\sigma_*(\ch(G)\op{Td}(B))$ giving
\begin{align*}
\ch_1(\sigma_*G)&=m\sigma\\
\ch_2(\sigma_*G)&=- \frac{mc_1\sigma}2
\end{align*}
Inserting these expressions into \ref{ind} we get
$$
I_X=- \frac12 mc_3(V)
$$
as we expected in \ref{lhs}. This gives us a simple test that our reduction
to the spectral data leads to the same result as it should be.

\section{Computation of $c_1({\F})$}

The first step is to provide some additional information about the
restriction of the spectral line
bundle $L$ to the intersection curve $S$! Following \cite{FMW} we
have
$$
c_1(L)= \frac12 (c_1(B)-c_1(C))+\gamma\,.
$$
Now let us restrict to $S=C\sigma$! As $C$ and $B$ are both divisors in
$X$ it follows from adjunction
formula \cite{GH} that
\begin{align*}
 c_1(C)_{\vert S} &= - C^2\sigma\\
c_1(B)_{\vert S}&=-C\sigma^2
\end{align*}

\begin{equation} c_1(L_{\vert S})= \frac12 (-\sigma^2C+\sigma
C^2)+\gamma_{\vert
S}
\label{rest}
\end{equation}
As mentioned before the condition $c_1(V)=0$ translates to fixing of
$\pi_*c_1(L)$ in $H^{1,1}(C)$ up to a class in $\ker
\pi_*\colon H^{1,1}(C)\rightarrow H^{1,1}(B)$ which is
$\gamma=\lambda(n\sigma-\eta+nc_1)$ as discussed in \cite{FMW},
so we get (note that $\lambda$ must be half-integral)
$$
\gamma_{\vert S}= -\lambda\eta C\sigma\,.
$$
In order to better understand \ref{rest} we will use a slightly different
perspective by computing $ i_*(c_1(L))$ in terms of the Chern classes
of $V$ using for instance (2.33) of \cite{OUR} taking into account that
$i_*L=\hat {\varPhi}  ^1(V)$. Let us write
$$
\ch_0(V)=n\,,\quad \ch_1(V)=0\,,\quad \ch_2(V)=-\eta\sigma+aF
$$
with  $\eta\in p^* H^2(B)$; we put the minus sign in $\ch_2(V)$ so that
$C=\ch_1(i_*L)=n\sigma+\eta$. Then,
$$
\ch_2(i_*L)= \frac12 nc_1(B)\sigma-(ch_3(V)- \frac12\eta
c_1(B)\sigma)\cdot F
$$
and Grothendieck Riemann-Roch theorem gives
$$
\ch_2(i_*L)=i_*(c_1(L)-\frac12 
c_1(N_{X/C}))=i_*(c_1(L))- \frac12 C^2
$$
so that
\begin{align*}
i_*(c_1(L))\sigma &= -\frac12 n c_1(B)^2\sigma+ \frac12
c_1(B)\eta\sigma+\frac12 C^2\sigma-\ch_3(V)
\\
& = \frac (C^2\sigma-C\sigma^2)-\frac12 c_3(V)
\end{align*}
Now we want to compute $c_1(L_{\vert S})\in H^2(S)$; if we understood
this
class as a number, this is the intersection number of the class
$c_1(L)\in
H^2(C)$ with the class of $S$ in $C$. Since $S=C\cdot\sigma$, we can
simply
compute the intersection number in $X$, thus obtaining $c_1(L_{\vert
S})=i_*(c_1(L))\sigma$ as numbers. It follows that
$c_3(V)/2=\lambda\eta(\eta-nc_1)\sigma$ which is agreement with results
in \cite{C} and \cite{DI}. So we obtain
\begin{equation}
c_1(L_{\vert S})=\frac12 (-\sigma^2C + \sigma C^2)-\frac12 c_3(V)\,.
\label{resst}
\end{equation}
Now we recall that $c_1(M)=0$ and $K_C={K_X}_{\vert C}+N_{X/C}=N_{X/C}$
such
that
${K_C}_{\vert S}=(N_{X/C})_{\vert S}$ and we find with \ref{resst}
$$
c_1({\F})= \frac12  m(3C\sigma^2+C^2\sigma) +  \frac12 m c_3(V)\,.
$$

\section{Index and Sections}

In this appendix we will show that
\begin{equation}
\begin{aligned}
h^0(S,L^{-1}_{\vert S}\otimes G_{\vert S})&=\frac12 mc_3(V)-mC^2\sigma \\
h^1(S,L^{-1}_{\vert S}\otimes G_{\vert S})&= -mC\sigma^2
\end{aligned}
\label{nos}
\end{equation}
Therefore let us consider the index
$$
I_X = - \dim Ext^1_C(L,\tilde\sigma_*(G_{\vert S})) -
\chi(S,{\F})
$$
and recall that $\chi(S,{\F})=mC\sigma^2+\frac12 mc_3(V)$ and
further we can write
$$
-\dim \Ext^1_C(L,\tilde\sigma_*(G_{\vert S}))=\chi(S,
L^{-1}_{\vert S}\otimes G_{\vert S}) -
h^0(S,L^{-1}_{\vert S}\otimes G_{\vert S})
$$
Applying the Riemann-Roch theorem we compute
$$
\chi(S, L^{-1}_{\vert S}\otimes G_{\vert
S})=m(C\sigma^2-C^2\sigma)+\frac12 mc_3(V)
$$
using the fact that
$$
c_1(L_{\vert S}^{-1}\otimes G_{\vert S})=\frac12 
m(3C\sigma^2-C^2\sigma) +  \frac12  m c_3(V)\,.
$$
Thus we find \ref{nos} from
$$
h^0(S,L^{-1}_{\vert S}\otimes G_{\vert S})=\chi(S, L^{-1}_{\vert
S}\otimes G_{\vert S})- \chi(S,{\F})-I_X.
$$

\end{document}